\documentclass[aps,prl,reprint,showpacs,superscriptaddress]{revtex4-1}

\usepackage{graphicx}
\usepackage{verbatim}
\usepackage{float}
\usepackage{sidecap}
\usepackage{dcolumn}
\usepackage{bm}
\usepackage{amssymb}
\usepackage{amsmath}	
\usepackage{color}
\usepackage{multirow}
\usepackage[a4paper]{geometry}
\newgeometry{top=1.5cm,right=1cm,left=1.5cm,bottom=1.5cm}
\usepackage[colorlinks=true, urlcolor=magenta, pdfborder={0 0 0}]{hyperref}
\hypersetup{
linkcolor=magenta,
citecolor=magenta
}

\newcommand{\spss}{$sp^{3}s^{*}$ }
\newcommand{\GaBiAs}{GaBi$_{x}$As$_{1-x}$ }
\newcommand{\GaBiAsGaAs}{GaBi$_{x}$As$_{1-x}$/GaAs }

\begin{document}

\title{Large-scale atomistic simulations demonstrate dominant alloy disorder effects in \GaBiAsGaAs multiple quantum wells}

\author{Muhammad Usman} \email{usman@alumni.purdue.edu} \affiliation{School of Physics, The University of Melbourne, Parkville, Melbourne, 3010, Victoria Australia.} 
\vskip 0.25cm

\begin{abstract}
Bismide semiconductor materials and heterostructures are considered a promising candidate for the design and implementation of photonic, thermoelectric, photovoltaic, and spintronic devices. This work presents a detailed theoretical study of the electronic and optical properties of strongly-coupled \GaBiAsGaAs multiple quantum well (MQW) structures. Based on a systematic set of large-scale atomistic tight-binding calculations, our results reveal that the impact of atomic-scale fluctuations in alloy composition is stronger than the inter-well coupling effect, and plays an important role in the electronic and optical properties of MQW structures. Independent of QW geometry parameters, alloy disorder leads to a strong confinement of charge carriers, a large broadening of the hole energies, and a red shift in the ground-state transition wavelength. Polarisation-resolved optical transition strengths exhibit a striking effect of disorder, where the inhomogeneous broadening could exceed an order of magnitude for MQWs, in comparison to a factor of about three for single quantum wells. The strong influence of alloy disorder effects persists when small variations in the size and composition of MQWs typically expected in a realistic experimental environment are considered. The presented results highlight the limited scope of continuum methods and emphasise on the need for large-scale atomistic approaches to design devices with tailored functionalities based on the novel properties of bismide materials.

\end{abstract}


\maketitle

\section*{1. INTRODUCTION}

Bismide alloys formed by dilute impurity concentrations of Bi atoms in GaAs are a rapidly emerging material system for the design of photonic devices~\cite{Bismuth_containing_compounds_2013, Wu_ACSP_2017, Marko_SR_2016, Ludewig_APL_2013, Janotti_PRB_2002}, photovoltaics~\cite{Richards_SEMSC_2017, Kim_JCG_2016, Johnson_patent_2014}, spintronic devices~\cite{Fluegel_PRL_2006}, and thermoelectric applications~\cite{Dongmo_JAP_2012}. The experimental and theoretical investigations of both unstrained and strained bulk \GaBiAs alloys have shown promising properties with increasing Bi fractions such as a large band gap reduction~\cite{Janotti_PRB_2002, Zhang_PRB_2005, Batool_JAP_2012, Usman_PRB_2011}, a crossover between the band gap and spin split-off energies~\cite{Batool_JAP_2012, Usman_PRB_2011, Usman_PRB_2013}, and the possibility of lattice-matched growth on a GaAs substrate~\cite{Broderick_SST_2013}. These novel characteristics have sparked a remarkable experimental interest in designing devices based on \GaBiAsGaAs quantum well (QW) structures, which could offer optimised performance~\cite{Richards_JCG_2015, Balanta_JoL_2017,Mazur_JoL_2017, Aziz_APL_2017, Makhloufi_NRL_2014, Marko_SR_2016, Balanta_JPD_2016, Patil_Nanotechnology_2017}. 

Despite significant ongoing experimental efforts~\cite{Richards_JCG_2015, Balanta_JoL_2017,Mazur_JoL_2017, Aziz_APL_2017, Makhloufi_NRL_2014, Marko_SR_2016, Balanta_JPD_2016, Patil_Nanotechnology_2017} on \GaBiAsGaAs MQW structures and supperlattices, the existing literature lacks theoretical guidance on the understanding and design of these nano-structures. To fully exploit the unique characteristics of \GaBiAsGaAs multiple quantum well (MQW) structures for a desired device operation, it is imperative to properly understand their properties in terms of geometry parameters and fluctuations in the spatial positions of Bi atoms (also known as alloy disorder). In particular, the interplay between the alloy disorder, inter-well coupling and geometry parameters has not been studied in the literature and it is yet unknown which of these three effects will dominate and dictate the overall electronic and optical properties of MQWs. This work, based on large-scale atomistic simulations, provides key insights pertaining the role of alloy disorder in \GaBiAsGaAs MQW structures and the established knowledge will provide useful guidance to the ongoing experiments for the design of future devices with tailored functionalities.  

\begin{figure*}
\includegraphics[scale=0.38]{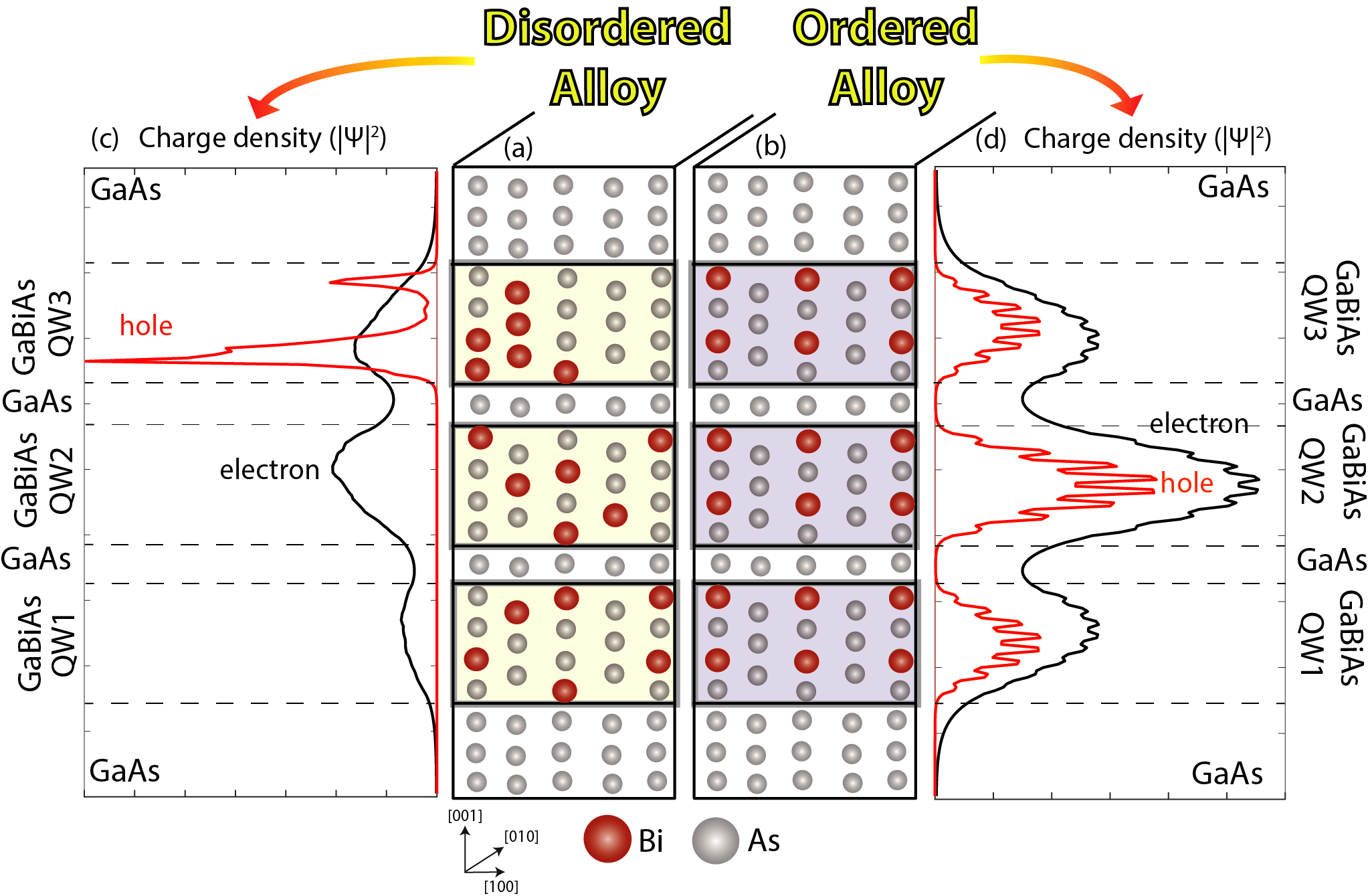}
\caption{\textbf{Ordered vs. disordered multiple quantum wells:} (a,b) Schematic diagram illustrates the position and type of atoms in a (010)-oriented anion-plane of a disordered and an ordered quantum well structure. Both MQW structures consists of three \GaBiAs quantum wells, separated by GaAs spacer layers. The geometry parameters and the Bi compositions of the two MQW structures are identical, except the spatial distribution of Bi atoms in the QW regions. The MQW structure in (a) has Bi atoms randomly distributed in the QW regions (disordered alloy), whereas the MQW in (b) exhibits uniform distribution of Bi atoms (ordered alloy). The impact of fluctuations in Bi atom positions is clearly evident in the electron and hole charge density plots shown in (c) and (d) for the disordered and ordered cases, respectively. The ordered MQW structure shows strongly coupled charge carrier states, whereas the disorder overcomes the inter-well coupling effect leading to much stronger confinement of charge carriers (in particular the hole state) due to the formation of Bi pairs and clusters.}
\label{fig:Fig1}
\end{figure*}

In this work, we have investigated the impact of alloy disorder by performing a systematic set of large-scale atomistic tight-binding simulations. We first start with hypothetical ordered QW devices, which contain ordered (uniform) distribution of Bi atoms such as illustrated in Fig.~\ref{fig:Fig1} (b). In the absence of disorder, the electronic properties of these devices are dictated by inter-well couplings. The confined charge carriers form strongly-coupled molecular states as evident from the electron and hole charge density plots shown in Fig.~\ref{fig:Fig1} (d). By increasing the number of QWs from one to four and also by increasing the separation between the QWs from 2 nm to 16 nm, we thoroughly study the evolution of electronic states governed by the underpinning coupling effect. Next, we switch on disorder in the QW regions by randomly placing the same fraction of Bi atoms as depicted in Fig.~\ref{fig:Fig1} (a). Keeping all other geometry parameters unchanged, a direct comparison with the ordered case reveals a strong impact of alloy disorder, which overcomes the coupling effect and dictates the confinement of charge carriers (Fig.~\ref{fig:Fig1} (c)). During the growth of QW devices, small uncertainties in the geometry parameters are inevitable. Therefore, we also study $\pm$1 nm variation in QW width and 0.5\% change in Bi composition. Our results indicate the dominant role of alloy disorder in spite of these fluctuations in the geometry parameters, confirming that the disorder related effects would control the electronic and optical properties of MQW structures under realistic experimental environments.   

The study of alloy disorder for a realistic \GaBiAsGaAs MQW device is a challenging problem because it requires theoretical modelling with atomistic resolution and the simulations need to be performed over large size of supercells~\cite{Usman_APL_2014, Usman_PRB_2013}. Majority of the existing literature on bulk \GaBiAs materials either rely on the simplified envelope wave function approximations such as effective-mass or k$\cdot$p~\cite{Marko_SR_2016, Broderick_SST_2015} which ignore alloy fluctuations, or based on DFT simulations~\cite{Polak_SST_2015, Bannow_arXiv_2017, Kudrawiec_JAP_2014, Bannow_PRB_2016} which are restricted to relatively small size of supercells due to the computational requirements. This work overcomes both challenges by establishing an atomistic tight-binding framework which can simulate realistic \GaBiAs MQWs with random placement of Bi atoms. Furthermore, the simulation domain in our work consists of several thousand atoms, capturing the true nature of disorder related effects.   

\section*{2. METHODS}

\subsection{2.1 Geometry Parameters}

Figure~\ref{fig:FigS1} shows the schematic diagram of a \GaBiAsGaAs MQW structure. In our study, a MQW device consists of $n$ number of QWs separated by $d$ nm thick GaAs spacer layers. To thoroughly investigate the effect of inter-well couplings, we have simulated MQW structures with $n \in$ \{1, 2, 3, 4\}. The separation $d$ between the QWs is selected as 4, 7, 11, 15, 21, and 29 monolayers (MLs), which is approximately 2, 4, 6, 8, 12, and 16 nm, respectively. The width $w$ of each QW region is 15 MLs ($\approx$ 8 nm). First, we have investigated inter-well couplings and alloy disorder effects based on uniform geometry parameters: $w$=8 nm and $x$=3.125\% for all the QW regions in a MQW device. In the last part of our work, we investigate the interplay between alloy disorder and geometry parameters by introducing $\pm$ 2 MLs (1 nm) variation in $w$ and 0.5\% variation in $x$.

For the uniform geometry MQW devices, 128 As atoms are replaced by Bi atoms to form \GaBiAs QWs with $x$=3.125\% composition. To focus on the effect of inter-well couplings, we first switch-off disorder in Bi atom positions and uniformly place 128 Bi atoms in each QW regions. For the disordered devices, 128 Bi atoms are placed randomly inside each QW region. The random distribution of Bi atoms is dependent on a seed value input to a random number generator that determines the nature of anion atom (either Bi or As) at a given atomic location inside the supercell. Different seed values ensure different arrangements of the Bi atoms, resulting in different numbers and types of Bi pairs and clusters in the QW region. It is also noted that a different seed value is used for each QW region within the same MQW device -- for example, for a double QW device ($n$=2), two different seed values were used for the two QW regions to ensure different configurations of Bi atoms. This set of two random distributions collectively is labelled as R1 to indicate a random configuration of the whole two quantum well device. For the computation of inhomogeneous broadening ($\Delta \lambda$) of electron and hole energies, we simulated five different sets of random configurations of Bi atoms (labelled as R1 to R5) for each MQW device.

It is important to simulate large size of supercell for proper modelling of the alloy disorder effects ~\cite{Usman_PRB_2013}. We have probed the electron and hole energies as well as the associated inhomogeneous broadening in those energies when the strained \GaBiAs supercell size is increased from 1000 atoms to 512000 atoms~\citep[][Section S1]{Usman_PRM_SM_2018}. Our calculations show that the small size of supercells ($<$ 4096 atoms) artificially modifies the electron/hole energies and enhance $\Delta \lambda$ values. By increasing the supercell size from 8000 atoms to 512000 atoms, we find that the electron and hole energies change by less than 1 meV and 10 meV respectively, whereas the value of $\Delta \lambda$ for the hole energies is roughly 27 meV in good agreement with the measured value of 31 meV~\cite{Usman_PRB_2013}. We therefore believe that the supercell size of 8000 atoms or more is expected to provide a reliable estimate of the properties of strained \GaBiAs alloys. In this work on MQWs, we have performed large-scale atomistic simulations consisting of several thousand atoms in the supercell, where each QW region consists of 8192 atoms. 

The thickness of both GaAs substrate and capping layer is 29 MLs ($\approx$ 16 nm) along the (001) direction, which is sufficient to allow proper strain relaxation. The overall size of the simulated supercell for the smallest MQW device ($d$=2 nm, $n$=2, $w$=8 nm) is 50$\times$4$\times$4 nm$^3$ consisting of 51,200 atoms, and for the largest device ($d$=16 nm, $n$=4, $w$=8 nm) is 112$\times$4$\times$4 nm$^3$ consisting of 114,688 atoms. The boundary conditions are periodic in all three dimensions. 

\begin{figure}
\includegraphics[scale=0.9]{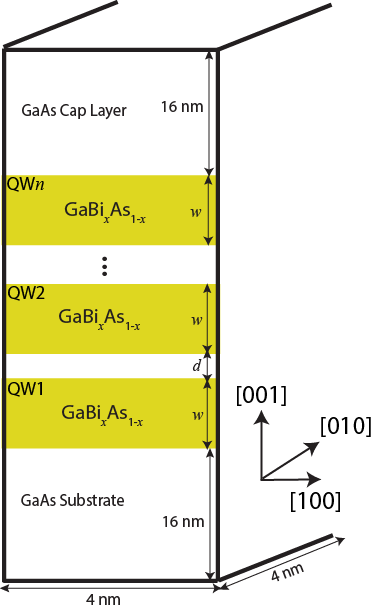}
\caption{Schematic diagram of a \GaBiAsGaAs multi-quantum-well structure is illustrated. The device consists of $n$ number of QWs with $d$ nm separation between the QWs and the width of each QWs is $w$ nm. The thickness of both GaAs substrate and capping layer is 16 nm, and the lateral dimensions are 4 nm in both (100) and (010) directions. The boundary conditions are periodic in all three spatial dimensions.}
\label{fig:FigS1}
\end{figure}

\subsection{2.2 Calculation of Strain Profiles}

In order to calculate strain induced by lattice-mismatch between GaAs and \GaBiAs materials, the \GaBiAsGaAs MQW supercells are relaxed by applying atomistic valence force field (VFF) energy minimization scheme~\cite{Keating_PR_1966, Lazarenkova_APL_2004, Usman_PRB_2011}. The VFF parameters for GaBi and GaAs materials are provided in Ref.~\cite{Usman_PRB_2011}. The values of $\alpha_{0}$ and $\beta_{0}$ for the GaAs are taken from Lazarenkova \textit{et al}. \cite{Lazarenkova_APL_2004}, whereas for GaBi are determined by obtaining relaxed bond lengths in accordance with Kent \textit{et al.}~\cite{Kent_PRB_2001}. 

After the VFF relaxation, the relaxed Ga-Bi bond-lengths are found to follow the trends of $x$-ray absorption spectroscopy measurements~\cite{Usman_PRB_2011}. The strain tensor components ($\epsilon_{xx}, \epsilon_{yy}, \epsilon_{zz}, \epsilon_{xy}, \epsilon_{yz}, \epsilon_{xz}$) are computed from the relaxed bond-lengths of atoms~\cite{Lazarenkova_APL_2004}. The strain parameters of interest, which are directly related to the energy shifts in the conduction and valence band edges are hydrostatic ($\epsilon_H$) and biaxial ($\epsilon_B$) strain components, which are defined as follows~\cite{Usman_PRB2_2011}: 
\noindent
\begin{eqnarray}
\epsilon_{\textrm H} &=& \epsilon_{xx} + \epsilon_{yy} + \epsilon_{zz} \\
\epsilon_{\textrm B} &=& 2 \epsilon_{zz} - \epsilon_{xx} + \epsilon_{yy}
\label{eq:strains}
\end{eqnarray}

It should be noted that for a compressively strained heterostructure such as studied in this work, the hydrostatic strain will be negative introducing a reduction in electron energy levels and the biaxial strain will be positive, leading to a net increase in the hole energy levels~\cite{Usman_PRB2_2011}.

\subsection{2.3 Calculation of Electronic Structure and Polarisation Resolved Optical Transition Strengths}

The electronic structure of MQW devices is computed from the nearest-neighbour ten-band \spss tight-binding (TB) theory, which explicitly include spin-orbit coupling. The TB parameters for GaAs and GaBi materials are derived~\cite{Usman_PRB_2011} to accurately reproduce the bulk band structure of these materials~\cite{Janotti_PRB_2002}. In the existing studies on both unstrained and strained GaBi$_{x}$As$_{1-x}$ alloys, the tight-binding model has shown good agreement with the available experimental data sets~\cite{Donmez_SST_2015, Balanta_JoL_2017, Zhang_JAP_2018, Dybala_APL_2017, Collar_AIPA_2017, Usman_PRB_2011, Usman_PRB_2013, Broderick_PRB_2014} as well as with the DFT calculations reported in the literature~\cite{Kudrawiec_JAP_2014, Polak_SST_2015, Bannow_PRB_2016}.

By solving the tight-binding Hamiltonian, we obtain ground state electron and hole energies and the corresponding wave functions at the $\Gamma$ point ($k$=0), which are labelled as $\vert \psi_{e} \rangle$ and $\vert \psi_{h} \rangle$, respectively, and are defined as: 
\noindent
\begin{eqnarray}
\vert \psi_{e} \rangle &=& \sum_{i,\mu} C^e_{i,\mu} \vert i\mu \rangle  \\
\vert \psi_{h} \rangle &=& \sum_{j,\nu} C^h_{j,\nu} \vert j\nu \rangle
\label{eq:e_h_wfs}
\end{eqnarray}
\noindent
where the label $i$ ($j$) represents the atom number inside the supercell and $\mu$ ($\nu$) denotes the orbital basis states on an atom for the electron (hole) states. $C^e_{i,\mu}$ and $C^h_{j,\nu}$ are the coefficients of the electron and hole wave functions, respectively, computed by diagonalising the TB Hamiltonian. 

The overlap between the electron and hole ground states is computed as follows:
\noindent
\begin{eqnarray}
\vert \langle \psi_{h} \vert \psi_{e} \rangle \vert &=& \vert  \sum_{i = j} \sum_{\mu = \nu} (C^h_{j,\nu})^* C^e_{i,\mu} \vert
\label{eq:e_h_overlap}
\end{eqnarray}
\noindent

The charge density plots of the electron and hole ground states as a function the distance along the (001) or $z$-axis are computed by calculating an average over the probability densities associated with all atom in the plane corresponding to each value at $z$-axis:
 \noindent
\begin{eqnarray}
\vert \psi_{e}(z) \vert^2 &=& \frac{1}{total \, \# \, of \, atoms \, at \, z}\sum_{all \, atoms \, at \, z} \vert \langle \psi_{e} \vert \psi_{e}  \rangle \vert^2 \\
\vert \psi_{h}(z) \vert^2 &=& \frac{1}{total \, \# \, of \, atoms \, at \, z}\sum_{all \, atoms \, at \, z} \vert \langle \psi_{h} \vert \psi_{h}  \rangle \vert^2
\label{eq:e_h_densities}
\end{eqnarray}

Note that we have only shown charge density plots for anion atom planes. The plots for cation atom planes exhibit similar behaviour~\cite{Usman_APL_2014} and are omitted for simplicity.

The inter-band momentum matrix elements between the ground electron and hole states is computed as follows~\cite{Usman_PRB2_2011}:
\noindent
\begin{eqnarray}
M_{\overrightarrow{n}}^{\alpha \beta} &=& \sum_{i,j} \sum_{\mu,\nu} (C^e_{i,\mu,\alpha})^* (C^h_{j,\nu,\beta}) {\langle i\mu\alpha \vert \textrm{\textbf{H}} \vert j\nu\beta \rangle} {(\overrightarrow{n}_{i}-\overrightarrow{n}_{j})} \label{eq:momentum_x}
\end{eqnarray}
\noindent
where $\alpha$ and $\beta$ represent spin of states, \textbf{H} is the \spss tight-binding Hamiltonian, and $\overrightarrow{n} = \overrightarrow{n}_{i} - \overrightarrow{n}_{j}$ is the real space displacement vector between atoms $i$  and $j$, and is either equal to  $\overrightarrow{x}_{i} - \overrightarrow{x}_{j}$ for the TE mode calculation or is equal to $\overrightarrow{z}_{i} - \overrightarrow{z}_{j}$ for the TM mode calculation. The optical transition strengths (TE$_{100}$ and TM$_{001}$) are then calculated by using Fermi's Golden rule and summing the absolute values of the momentum matrix elements over the spin degenerate states:
\noindent
\begin{eqnarray}
\textrm{TE$_{100}$} &=&  \sum_{\alpha, \beta} \vert M_{\overrightarrow{x}}^{\alpha \beta} \vert ^2 \label{eq:TE_X} \\
\textrm{TM$_{001}$} &=&  \sum_{\alpha, \beta} \vert M_{\overrightarrow{z}}^{\alpha \beta} \vert ^2  \label{eq:TM_Z}
\end{eqnarray}   

The tight-binding model is implemented with in the framework of atomistic tool NanoElectronic Modeling (NEMO 3-D) simulator~\cite{Klimeck_IEEETED_2007_1, Klimeck_IEEETED_2007_2} which has, in the past, shown an unprecedented accuracy to match experiments for the study of nano-materials~\cite{Usman_PRB_2011, Usman_PRB_2013, Usman_NN_2016} and devices~\cite{Usman_PRB2_2011, Usman_IOP_2012}. 

\section*{3. RESULTS AND DISCUSSIONS}

\subsection{3.1 Effect of inter-well coupling in ordered MQWs}

\begin{figure*}
\includegraphics[scale=0.38]{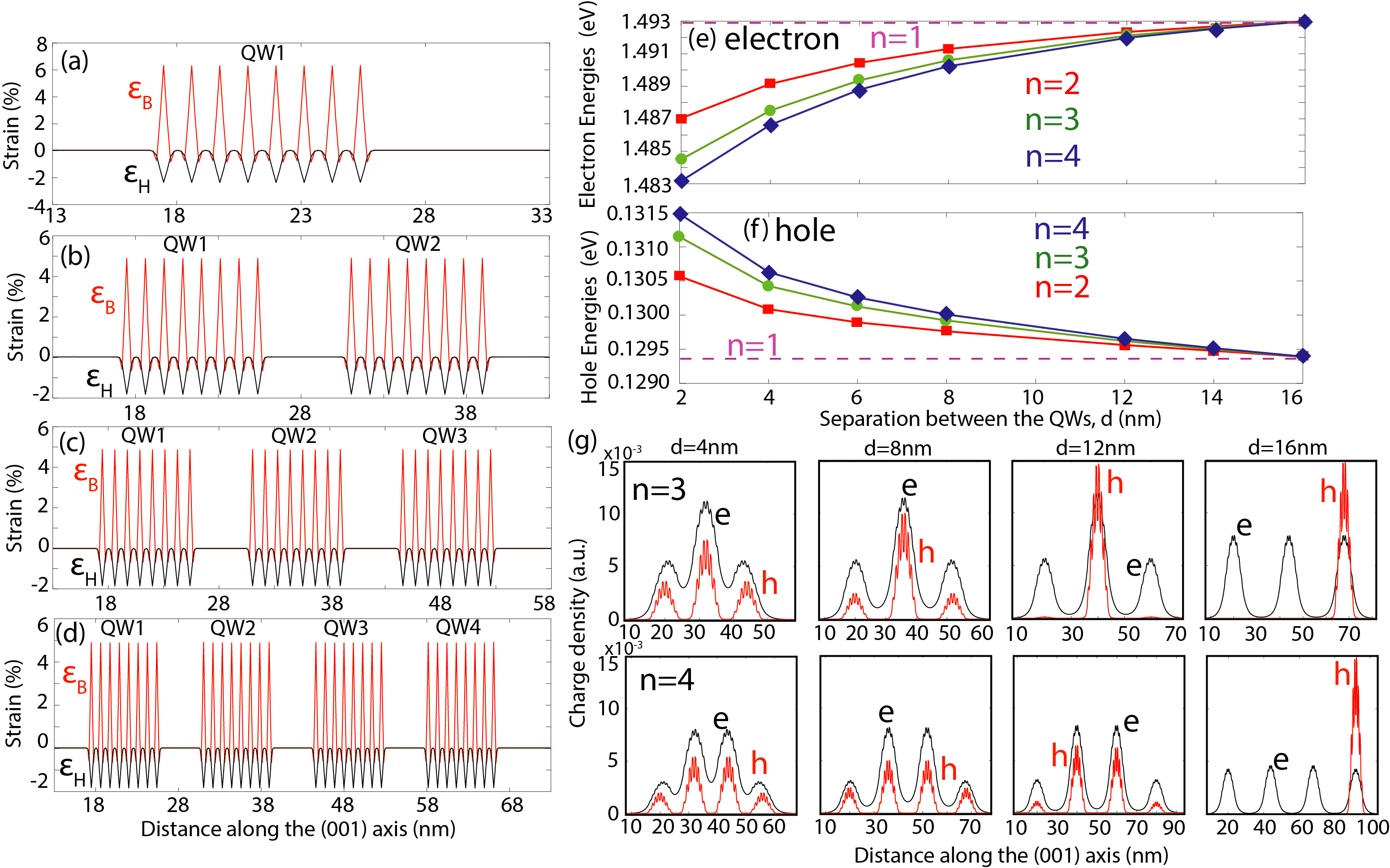}
\caption{\textbf{Inter-well coupling effects in ordered MQW devices:} (a-d) The line plots of hydrostatic ($\epsilon_H$) and biaxial ($\epsilon_B$) strain components as a function of distance along the (001) axis are shown through the centre of QWs for $n$=1, 2, 3, and 4. The separation $d$ between the QWs is 4 nm and the widths of QWs are 8 nm. (e, f) Plot of the lowest electron and the highest hole energy levels are shown as a function of the separation ($d$) between the QWs for $n$=2, 3, and 4. The horizontal dashed lines indicate energy levels for a single QW device. (g) Plots of the lowest electron and the highest hole state charge densities as a function of distance along the (001) axis are shown for a few selected separations $d$ and for $n$=3 and 4.}
\label{fig:Fig2}
\end{figure*}

To investigate the inter-well coupling effect, we first artificially switch off alloy disorder by uniformly placing 128 Bi atoms in the QW regions (one Bi atom per four unit cells of GaAs). Fig.~\ref{fig:Fig2} (a-d) plots the strain profiles for the ordered QW devices along the (001) axis through the centre of QWs for $n$=1, 2, 3, and 4 respectively, and $d$=4 nm. The peaks indicate the accumulation of strain in the vicinity of Bi atoms. The positive (negative) sign of $\epsilon_B$ ($\epsilon_H$) is due to the larger lattice constant of GaBi in comparison with GaAs, which is consistent with an existing study of III-V quantum dot heterostructure~\cite{Usman_PRB2_2011}. Overall we find that there is a small change in the hydrostatic and biaxial strain parameters when the number of QWs is increased. 

Fig.~\ref{fig:Fig2} (e) and (f) plots the lowest electron and the highest hole energies as a function of the QW separation for $n$= 1, 2, 3, and 4. As the separation between the QWs is reduced, the strong inter-well coupling leads to a decrease (increase) in the electron (hole) energies. To further study the strength of inter-well couplings as a function of $d$, we also plot line-cut profile of electron and hole charge (or probability) densities (computed by equations 6 and 7) as a function of distance along the (001) axis for a few selected inter-well separations. Fig.~\ref{fig:Fig2} (g) shows charge density plots for $n$=3 and 4, and the plots for $n$=1 and 2 are provided in the Ref.~\citep[][Section S2 and S3]{Usman_PRM_SM_2018}. The strong inter-well coupling leads to hybridized bonding-like electron and hole states for the small inter-well separations ($d$=4 and 8 nm). The coupling effect becomes gradually weak as the separation between the QWs is increased from 12 nm to 16 nm. In particular, in Fig~\ref{fig:Fig2} (g), the hole state is found to be completely de-coupled at 16 nm separation and is confined in the top-most QW region. This is due to the heavier effective-mass of holes which leads to stronger confinement effect, also previously observed in coupled quantum dots~\cite{Jaskolski_PRB_2006}. The electron states on the other hand experience relatively weaker confinement effect, and therefore become decoupled at around 24 nm separation. The overall shape and confinement of the electron and hole ground states for the ordered QW devices is consistent with what we expect from an envelope wave function approximation approach.  

\subsection{3.2 Disorder effects in realistic MQW structures}

In realistic QW devices, random fluctuations in the spatial positions of Bi atoms inside the \GaBiAs QW regions are expected and the recent experiments have shown the evidence of clustering of Bi atoms~\cite{Gogineni_APL_2013, Reyes_SST_2013}. Therefore it is crucial to investigate the effect of disorder to properly understand the experimental measurements and to design devices with desired functions. Here we study the effect of alloy disorder by randomly placing 128 Bi atoms ($x$=3.125\%) in each QW region of a MQW structure. To properly model the disorder in Bi atom positions and to quantify the related inhomogeneous broadening of electron and hole energies, we consider five different sets of random placements for Bi atoms as defined earlier in the section 2.1. 

\begin{figure*}
\includegraphics[scale=0.36]{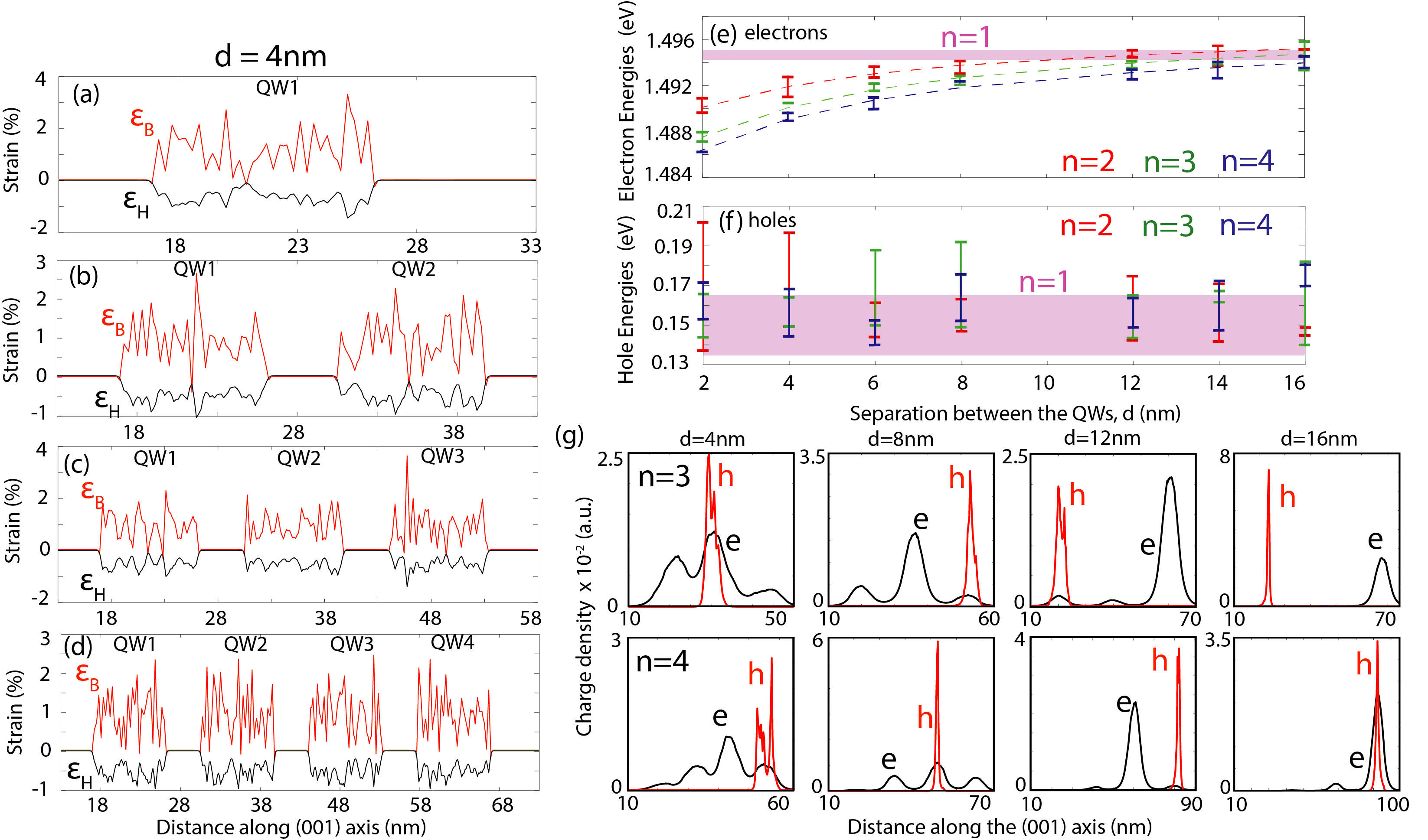}
\caption{\textbf{Dominant role of alloy disorder in MQW structures:} (a-d) The line plots of hydrostatic ($\epsilon_H$) and biaxial ($\epsilon_B$) strain components as a function of distance along the (001) axis are shown through the centre of QWs for $n$=1, 2, 3, and 4. The separation between the QWs is 4 nm and the widths of QWs are 8 nm. (e, f) Plots of the lowest electron and the highest hole energy levels are shown as a function of the separation ($d$) between the QWs for $n$=2, 3, and 4. The error bars shows the broadening of energies computed over five different random distributions of Bi atoms in the QW regions. The dashed lines in (e) are plotted as a guide to eye to indicate the overall trend of variation in electron energies. The shaded areas indicate energy level broadening for a single QW ($n$=1). (g) Plots of the lowest electron and the highest hole state charge densities as a function of distance along the (001) axis plotted are shown for a few selected separations $d$ and for $n$=3 and 4.}
\label{fig:Fig3}
\end{figure*}

Fig.~\ref{fig:Fig3} (a-d) plots the strain profiles for one particular exemplary random distribution of Bi atoms, and for $d$=4 nm and $n$=1, 2, 3, and 4 (the plots for other Bi configurations exhibit similar profile). The distribution of strain peaks is arbitrary due to the random occurrence of Bi atoms at the As positions and clearly emulate the disordered character. Notably compared to the strain profiles for the ordered cases, the overall magnitude of the hydrostatic and biaxial strain components is reduced by approximately a factor of two. This is because for the ordered cases, each Bi atom is connected to four neighbouring As atoms and therefore exhibits large peaks in the strain profiles with the magnitude of strain components reducing to zero in-between the peaks. The disordered alloy is a mixed configuration of Bi atoms including pairs and clusters of Bi atoms. Therefore, the strain profiles exhibit smaller peaks but more distributed nature, and do not become zeros in-between the peaks. Overall we expect a stronger effect of strain on the electronic properties of MQWs in the presence of disorder, as will be shown later in this section where larger shifts in the electron and hole energies are computed for disordered cases compared to the corresponding ordered case. 

Fig.~\ref{fig:Fig3} (e) and (f) plots the lowest electron and the highest hole energies as a function of the QW separations ($d$) for $n$=1, 2, 3, and 4. Both, the electron and the hole energies, are broadened due to the presence of alloy disorder. The broadening ($\Delta \lambda$) of electron energies is very small (of the order of 1 to 2 meV), which is also nearly independent of $n$ and $d$. The hole energies are significantly affected and show a much larger value of $\Delta \lambda$. The magnitude of $\Delta \lambda$ for the hole energies is further highlighted in the Ref. ~\citep[][Figure S4]{Usman_PRM_SM_2018} for $n$=1, 2, 3, and 4 devices. Our results indicate that for the single QW devices, $\Delta \lambda$ is about 31 meV, which varies over a wide range (from 4 meV to 64 meV) for the studied MQW structures ($n >$ 1). On average, we predict that a value of 10-40 meV is expected for $\Delta \lambda$ in the case of strongly-coupled MQW devices. The stronger disorder effect on the hole energies compared to the electron energies is expected because the Bi-related resonant states are predicted to lie closer to the valence band edge of GaAs, and therefore a strong band-anticrossing type interaction with the valence band edge will induce a correspondingly large disorder-related effect on the hole states~\cite{Usman_APL_2014, Usman_PRB_2011}. 

In terms of electron and hole energy shifts, we find that the dependence of the electron energies on $d$ is quite similar to what we computed for strongly coupled ordered QWs (decrease in energy when $d$ is reduced), which again indicates that the disorder related effects are relatively weak for the electron states. When the number of QWs are increased from 1 to 4, the overall decrease in the electron energies is around 10 meV, which is same as observed earlier for the ordered QW devices (Fig.~\ref{fig:Fig2} (e)). Disorder further increases the electron energies by about 1 meV. The hole energies on the other hand show large variations irrespective of $d$ and $n$, and exhibit no coupling related trend. Therefore, we cannot characterize the change in the hole energies as increase or decrease as a function of $d$. The introduction of alloy disorder however causes a large upward shift of about 60 meV in the hole energies. Our results predict a net red shift in the ground state transition energies due to the alloy disorder in the MQW structures. Finally for all the MQW structures investigated in this work, we compute an overall band gap reduction as 53-74 meV per \% Bi, which is in good agreement with the reported experimental values of 84-53 meV per \% Bi~\cite{Kudrawiec_JAP_2014, Ludewig_JCG_2013}.  

\begin{figure}
\includegraphics[scale=0.4]{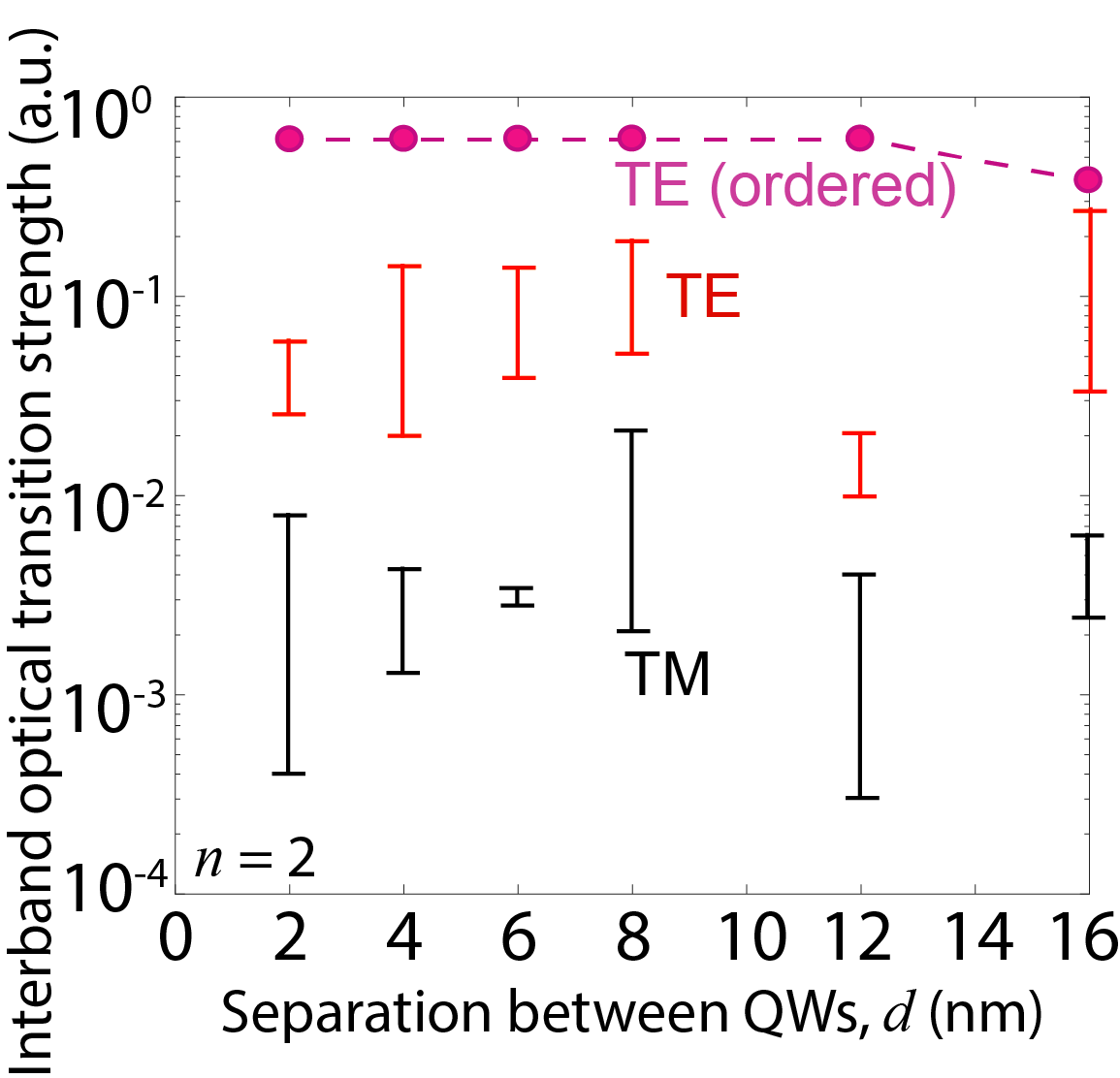}
\caption{\textbf{Polarisation-resolved inter-band optical transition strengths:} Plots of the polarisation-dependent inter-band optical transition strengths are shown as a function of the distance along the (001) axis for $n$=2. The error bars are for disordered \GaBiAs QW structures and indicate the broadening of transition strengths due to fluctuations in Bi atom positions.}
\label{fig:Fig4}
\end{figure} 

Although the variation in the electron and hole energies provides some understanding of the impact of disorder, the electron and hole charge density plots exhibit much more clear picture of the overall electronic characteristics of MQW devices. Fig.~\ref{fig:Fig3} (g) shows the charge density plots for a few selected values of $d$ and $n$ and one particular random distribution of Bi atoms. Plots for other values of $n$ and three sets of random distributions of Bi atoms (R1, R2, and R3) are provided in Ref.~\citep[][Section S4]{Usman_PRM_SM_2018}. For reference, we also provide plots for the single QW devices in Ref.~\citep[][Figure S2]{Usman_PRM_SM_2018}. The plots for the single QWs show that the hole states are significantly perturbed (much more confined), whereas the electron states largely retain their envelope wave function approximation type character. Strikingly for the coupled QW systems, we find that the electron states are also significantly perturbed and for some cases, their confinement is much stronger than what we expect solely based on the coupling effects. For example, in Ref.~\citep[][Figure S6]{Usman_PRM_SM_2018} the confinement of the electron states for $d$=6 nm and $n$=3 is entirely in the upper two QW regions (for all three sets of random distributions of Bi atoms) indicating the breakdown of coupling effect. It is not possible to capture such character of electron states in simple theoretical models such as effective-mass or k$\cdot$p, which ignore disorder. 

A second consequence of the alloy disorder effects on the electron states is evident from the plots for $d$=16 nm case, where the electron states show single QW confinements in contrast to the ordered case where such confinement is possible at much larger values of $d$. Therefore we show that the existing notion, based on bulk alloy studies and single QW structures, that the alloy disorder negligibly impacts the probability density of conduction band states cannot be applied to MQW structures.   

The hole states do not exhibit any coupling effect in the presence of disorder and are always strongly confined in a single QW region for all of the investigated MQW structures. Moreover, for the same Bi composition and QW geometry parameters, we find that the hole wave functions are confined in different QW regions just based on the variation in the atomic configurations of Bi atoms (see for example d=2 nm in \citep[][Figure S7]{Usman_PRM_SM_2018}). Based on our previous study~\cite{Usman_PRB_2011} which demonstrated larger valence-band energy shifts and stronger hole state confinement in the presence of Bi pairs and clusters, we infer that the strength of disorder is stronger in the QW regions where the hole wave functions are confined. This arbitrary nature of hole confinements implies that the electron-hole wave function overlap would significantly vary from device to device even with the same set of geometry parameters: \{$x$, $n$, $d$, and $w$\}. This character is shown in the Ref.~\citep[][Figure S8]{Usman_PRM_SM_2018}, where we have plotted electron-hole wave function overlaps as a function of $d$ for $n$=2, 3, and 4. For reference, we have also included values for the corresponding ordered cases. The ordered cases show an almost constant value of overlap for small values of $d$ (strongly coupled regime), and a decrease as $d$ becomes large. However, for the disordered cases, not only a significant reduction in the electron-hole wave function overlap is computed due to much stronger confinement of the hole states, but also there is a large variation in the magnitudes of the computed overlaps due to an arbitrary nature of the hole positioning inside the QW devices. 
 
\subsection{3.3 Polarisation-resolved optical transitions strengths}

Following parameters contribute to the polarisation-dependent TE and TM inter-band optical transition strengths between an electron and a hole states: (i) the overlap between the electron and hole wave functions, (ii) the coupling between QWs which increases TM mode, and (iii) the conventional selection rule which allows transitions between Eigen states of certain character. For a compressively strained QW such as \GaBiAsGaAs, the highest hole state will have heavy-hole type character and therefore TM mode transition is prohibited based on the selection rule. Indeed we find zero magnitude of TM mode transition for the ordered QW structures. The presence of disorder breaks down the coupling effect and the selection rule, allowing some mixing of $p_z$ (light-hole) character in the highest hole state~\cite{Usman_PRB_2013}, giving rise to a small amplitude of TM mode transition as shown in Fig.~\ref{fig:Fig4}.  

\begin{figure*}
\includegraphics[scale=0.4]{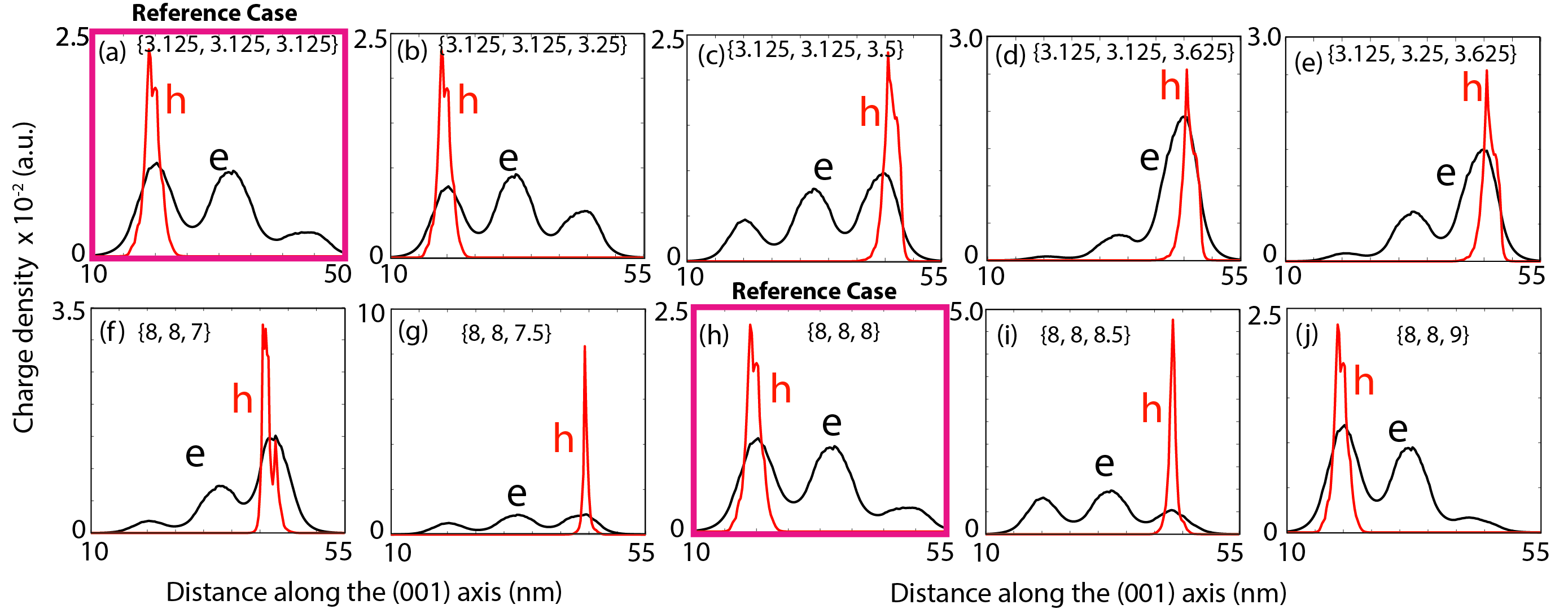}
\caption{\textbf{Interplay between alloy disorder and geometry parameters:} (a-e) The Bi fraction of a triple QW structure ($n$=3) is varied as indicated by labels \{$x_1,x_2,x_3$\}, where $x_1$,$x_2$, and $x_3$ are the Bi fractions of the lowest, middle and top QWs, respectively. The widths $w$ of all three QW regions are 8 nm and the separation between the QWs is 4 nm. (f-j) The widths of QWs in a triple QW structure ($n$=3) are varied as indicated by labels \{$w_1,w_2,w_3$\}, where $w_1$,$w_2$, and $w_3$ are the widths of the lowest, middle and top QWs, respectively. The Bi fraction $x$ of all three QW regions is 3.125\% and the separation between the QWs is 4 nm.}
\label{fig:Fig5}
\end{figure*} 

The magnitude of the TE mode transition is strongly dependent on overlap between the electron and hole wave functions. As we have discussed in the previous sections that the presence of disorder significantly perturbs the electron and hole states, leading to a strong confinement of hole states, which reduces electron hole wave function overlaps. Moreover the overlap between the electron and hole states exhibits a large variation. These two effects will not only reduce the overall strength of TE mode, but also cause a large variation in its magnitude. This is shown in Fig.~\ref{fig:Fig4} where we have plotted TE mode strengths as a function of the QW separations for $n$=2. Same plots for $n$=3 and 4 are provided in Ref.~\citep[][Figure S9]{Usman_PRM_SM_2018}. For comparison, we also plot the corresponding values for the ordered cases. A clear reduction in the inter-band transition strengths is observed for the disordered structures. Large variations in TE mode are a direct consequence of highly confined hole states which lead to large variations in the wave function overlaps. Notably, we find that the spread of the TE mode variation is an order of magnitude, which is significantly higher than the single QW devices where the extent of this spread was about a factor of three~\cite{Usman_APL_2014}. This is due to the fact that for single QW device, electron wave function is negligibly perturbed and retains its envelope wave function approximation type character. Moreover both electron and hole states reside in the same QW region, therefore relatively small changes occur in the electron-hole wave function overlap. Contrarily for MQW structures, the stronger confinement of both electron and hole states, in conjunction with the arbitrary selection of a QW region for the hole confinement leads to a much larger variation in the inter-band transition strengths. It is also noted that we have only considered the lowest electron and the highest hole states in the calculations of TE and TM modes. However due to the breakdown of selection rule for disordered alloys, non-zero magnitude of TE mode transitions is expected between the lowest electron state and a few highest hole states~\cite{Usman_APL_2014}, which will further increase the broadening of TE mode transition energy and strength. 

\subsection{3.4 Interplay between disorder and geometry parameters}

From Fig.~\ref{fig:Fig3}, it is clear that the presence of disorder strongly influences the electronic properties of MQW devices. However in our discussion so far, we have fixed the widths ($w$=8 nm) and Bi compositions ($x$=3.125\%) of QWs. In a realistic experimental environment, these two geometry parameters are hard to control, and small variations are always observed. Therefore, it is important to investigate the alloy disorder effects in the presence of small variations in $w$ and $x$. 

Let's first investigate the Bi composition of QWs. For this purpose, we select an exemplary triple QW structure with $d$=4 nm and $w$=8 nm. The Bi compositions of three QWs in this device are represented by \{$x_1,x_2,x_3$\}. Fig.~\ref{fig:Fig5} (a) plots the electron and hole charge densities for the reference case, where the compositions are uniform ($x_1=x_2=x_3$=3.125\%). The confinement of hole state in the lowest QW region indicates that the disorder is strongest in this QW region. It is known from the composition-dependence of electron and hole energies in bulk \GaBiAs alloys that increasing $x$ of a QW will increase (decrease) the valence (conduction) band energies for that particular QW~\cite{Usman_PRB_2011}, thereby pulling the electron and hole states inside it. In Figs.~\ref{fig:Fig5} (b) to (d), we gradually increase the Bi composition of the topmost QW and plot the electron and hole charge densities. The electron charge density gradually moves towards the upper QWs, however the hole charge density remains confined in the lowest QW until $x$ is increased by 0.25\% or more. This shows that the alloy disorder effect is dominant over the Bi composition effect for up to 0.25\% variation in $x$. The same results are obtained for a second case of a quadruple MQW structure discussed in Ref.~\citep[][Section S5]{Usman_PRM_SM_2018}. Note that we have chosen the extreme case by increasing $x$ of the topmost QW region, when the hole state was confined in the lowest QW region. If instead $x$ for the lowest QW is increased, the hole state will stay confined inside it and the electron state will also increasingly confine in the lowest QW. The same effect is shown in Fig.~\ref{fig:Fig5} (e), where we have increased the Bi composition of the middle QW region compared to the part (d). The hole state remains confined in the upper-most QW indicating a strong disorder effect, but the confinement of the electron state slightly increases in the middle QW region. 

Next, we investigate the fluctuations in the width of QW ($w$) by keeping the other parameters fixed at $x$=3.125\%, $d$=4 nm. Fig.~\ref{fig:Fig5} (h) shows the reference case where all the three QWs have a uniform width of 8 nm. In conventional MQW structures, if the width of a QW region is increased (decreased), it reduces (increases) the confinement energies of charge carriers and therefore the ground electron and hole states will tend to reside in the larger width QW region~\cite{Hall_NJP_2013}. However for highly mismatched alloys such as \GaBiAs, if $x$ is kept fixed, the increase (decrease) in the QW width will lead to a weaker (stronger) alloy disorder effect and therefore it will push the electron and hole ground states away from the larger QW region. Therefore changing the width of a QW introduces two competing effects (from disorder and from the change in confinement energies). This interplay between the two competing effects is observed in Figs.~\ref{fig:Fig5} (f-j). When the QW width is decreased from 8 nm to 7 nm, both the electron and hole states moves into the smaller width QW indicating alloy disorder effect dominating the QW width effect. On the other hand, for the larger QW width (changing from 8 to 9 nm), at first the geometry effect pushes the hole state towards the wider QW region, but at 9 nm the alloy disorder effect again dominates and leads to the electron and hole states being outside of the wider QW region. In Ref.~\citep[][Section S5]{Usman_PRM_SM_2018}, we have discussed the impact of small variations in $w$ for a second case with $n$=4, concluding that the alloy disorder effects overall dominate the impact of QW width variations. 

\section*{4. CONCLUCIONS}

Based on large-scale atomistic tight-binding calculations, we have shown a strong influence of the alloy disorder effects on the electronic and optical properties of \GaBiAsGaAs MQW devices. The direct comparison of ordered and disordered QW structures provides a quantitative estimate of the disorder related energy shifts in the electron and hole ground states as well as the character of their confinements. We show that the electron states in MQWs are also perturbed by the fluctuations in Bi atom positions, in contrast to the existing studies for bulk and single QWs which suggest that the conduction band states are negligibly impacted by disorder. For the studied disordered MQW structures, we compute a ban-gap reduction of 53-75 meV per \% Bi, which is in good agreement with the reported experimental range of 53-84 meV per \% Bi. Based on five sets of random configurations of Bi atoms for each MQW device, we estimate a 10-40 meV broadening of the highest hole state energy and an order of magnitude broadening of the TE mode transition strength between the lowest electron state and the highest hole state. We also consider small variations in the geometry parameters of MQW structures consistent with the experimental growth uncertainties, and conclude that the effects of disorder are independent of such variations. The presented results provide critical insights for the understanding and designing of MQW devices based on novel characteristics of bismide materials and will be useful for a wide range of applications in photonics, photovoltaics, thermoelectric areas of research. 

\textbf{\textit{Acknowledgements:}} Computational resources are acknowledged from NCN/Nanohub. Some parts of this work were carried out at Tyndall National Institute, Cork Ireland. 
\\ \\

\clearpage
\newpage

\textbf{\textit{\underline{Supplementary Material Section}}}
\\ \\

\renewcommand{\thefigure}{S\arabic{figure}}
\setcounter{figure}{0}

\renewcommand{\theequation}{S\arabic{equation}}
\setcounter{equation}{0}

\noindent
\\ \\
\textbf{S1. The effect of supercell size on electron and hole energies}
\\ \\
The size of supercell is an important parameter to accurately model the effects of alloy disorder. In our previous study~\cite{Usman_PRB_2013}, we have probed supercell size effect on the character of hole states, which indicated that a supercell size containing 4096 or more atoms is crucial to understand the alloy properties. Here we extend that study to larger size of supercells containing up to 512000 atoms and investigate the variation in the lowest electron and the highest hole energies. The supercell contained 3.125\% Bi atoms. We also simulate five different random configuration of Bi atoms for each supercell size and compute inhomogeneous broadening ($\Delta \lambda$) of the electron and hole energies. The computed results are plotted in Fig.~\ref{fig:FigS1}. Our calculations indicate that when the supercell size is increased from 8000 atoms to 512000, the electron energies and the associated broadening only negligibly vary. On the other hand, the variation in the hole energies is approximately within 10 meV. The broadening of hole energies is also computed as approximately 27 meV which is consistent with supercell size and is also in good agreement with the measured value~\cite{Batool_JAP_2012, Usman_PRB_2013}. Based on our results, we conclude that a supercell size of 8000 atoms captures the alloy properties with reasonably good accuracy without demanding extensive computational resources.    

\noindent
\\ \\
\textbf{S2. Charge densities for single quantum well devices}
\\ \\
Figure~\ref{fig:FigS2} plots the charge densities for single QW devices as a function of the distance along the (001) axis. For a direct comparison, we show both ordered and disordered (three different random configurations of Bi atoms) charge density plots. From the ordered QW simulation, the plots of charge density closely resemble to what we expect from envelope wave function approximation methods. Both electron and hole states are symmetrically confined, with hole confinement being slightly stronger due to the associated heavier mass. 

When the alloy disorder is introduced in the QW regions, the electron wave function is weakly affected and retains some characteristic of the envelope-type approximation. However the hole wave function exhibits a strong perturbation and is heavily confined in the QW region. This is expected because the Bi related resonance states are predicted to lie below the valence band edge of the GaAs material and strongly interact with the valence band through band-anticrossing interaction~\cite{Usman_PRB_2011, Broderick_SST_2012}. Therefore the effect of Bi pairs and clusters will strongly modify the hole energies and the confinement of hole states.   

\noindent
\\ \\
\textbf{S3. Charge densities for ordered double quantum well structures}
\\ \\
Charge density plots for the triple and quadruple QW structures with ordered distribution of Bi atoms in the QW regions are shown in Fig. 2 of the main text. Here we show the same plots for double QW structures. Overall the character of the electron and hole ground states is consistent with what we expect from strongly coupled MQWs. At small values of QW separation, the two QWs are strongly coupled and therefore the electron and hole states form hybridized molecular states. For large separation, the coupling between the QWs becomes weak and therefore the electron and hole states tend to confine in individual QW regions. We find that coupling vanishes much quickly for the holes states and at $d$=16 nm, the ground hole state is nearly entirely confined in a single QW region. On the other hand, the electron state is still weakly coupled even at 16 nm and turn into a completely decoupled state at about 24 nm separation.

\noindent
\\ \\
\textbf{S4. Charge densities for disordered MQW structures}
\\ \\
The confinement of electron and hole wave functions in disordered MQW structures is strongly modified by the presence Bi pairs and clusters. Therefore it is expected that the charge density plots will exhibit drastically different confinements as a function of fluctuations in Bi positions. To demonstrate this effect, in this section we have considered three different random configuration of Bi atoms for each MQW structure and plotted the corresponding electron and hole charge densities as a function of the distance along the (001) axis. The random Bi configurations are labelled by $R_1$, $R_2$, and $R_3$. Figs.~\ref{fig:FigS5}, ~\ref{fig:FigS6}, and ~\ref{fig:FigS7} show plots for MQW structures with $n$=2, 3, and 4, respectively. The charge density plots clearly show the strength and nature of alloy disorder effects on the character of electron and hole states in MQW devices.  Overall we deduce the following important insights which are expected to be critical to understand future experimental measurements:
\\ \\
\noindent
1) The inter-well coupling effect is completely vanished for the hole states, which are strongly confined in only one QW region for all values of $n$ and $d$. Furthermore, the selection of a particular QW region for hole confinement is entirely arbitrary and is dependent on strength of disorder for that particular QW region.   
\\ \\
\noindent
2) The electron states still exhibit inter-well coupling effects but this effect is significantly weakened by the disorder. For 
\\ \\
\noindent
3) From the ordered MQW structures, we have found that the inter-well coupling for electron states vanishes for $d$=24 nm. However due to disorder related weakening of coupling, typically at 16 nm separation we find that the electron states are confined in individual QW regions and therefore exhibit no coupling effect. 

\noindent
\\ \\
\textbf{S5. Interplay between alloy disorder and geometry parameters}
\\ \\
In the main text of the paper, we have investigated the role of alloy disorder in the presence of small fluctuations in Bi composition $x$ and width of QW $w$ for a three quantum well structure with 4 nm separation between the QW regions. In this section, we study a second case where similar changes in $x$ and $w$ are introduced for a four QW structure with 4 nm separation between the QWs. 

Let us first investigate the effect of Bi composition. Fig.~\ref{fig:FigS7} (a) shows the plot of electron and hole charge densities for reference case when all the four QWs have same 3.125\% of Bi atoms. The alloy disorder dictates the confinement of hole state in the upper most QW region where as the electron state is weakly hybridized and mainly present in the upper two QWs. To check the extreme case, we increase the Bi composition of the lowest QW region in (b) and (c). The increased Bi composition of the lowest QW means it should now host the electron and hole states. However our calculations indicate that the hole state remains confined in the upper most QW region thus implying a strong alloy disorder effect which overcomes the Bi composition effect. The electron state however is slightly shifted towards the lower QW region as it experiences weaker impact from the alloy disorder.    

In Figs.~\ref{fig:FigS7} (d-f), we investigate the effect of QW width variations. In part (e), we plot the reference case where all of the four QWs have uniform (8 nm) widths. When the width of a QW region is decreased (increased), the electron and hole states should move out (inside) that QW region. However in part (d), the hole state is moved inside the lowest QW region despite its smaller width of 7 nm. This is because the smaller width of QW would lead to a larger number of Bi pairs and clusters thus increasing the strength of alloy disorder which in spite of QW width induced energy shift confines the hole state in smaller width QW. On the other hand, when the lowest QW width is increased to 9 nm (in part f), the hole state is still confined in the uppermost QW region and the small change in hole energy is not sufficient to overcome the larger change from the alloy disorder. Overall these results clearly show that the alloy disorder is dominant over the $\pm$1 nm fluctuations in the QW width.   

\newpage
\clearpage

\begin{figure*}
\includegraphics[scale=0.8]{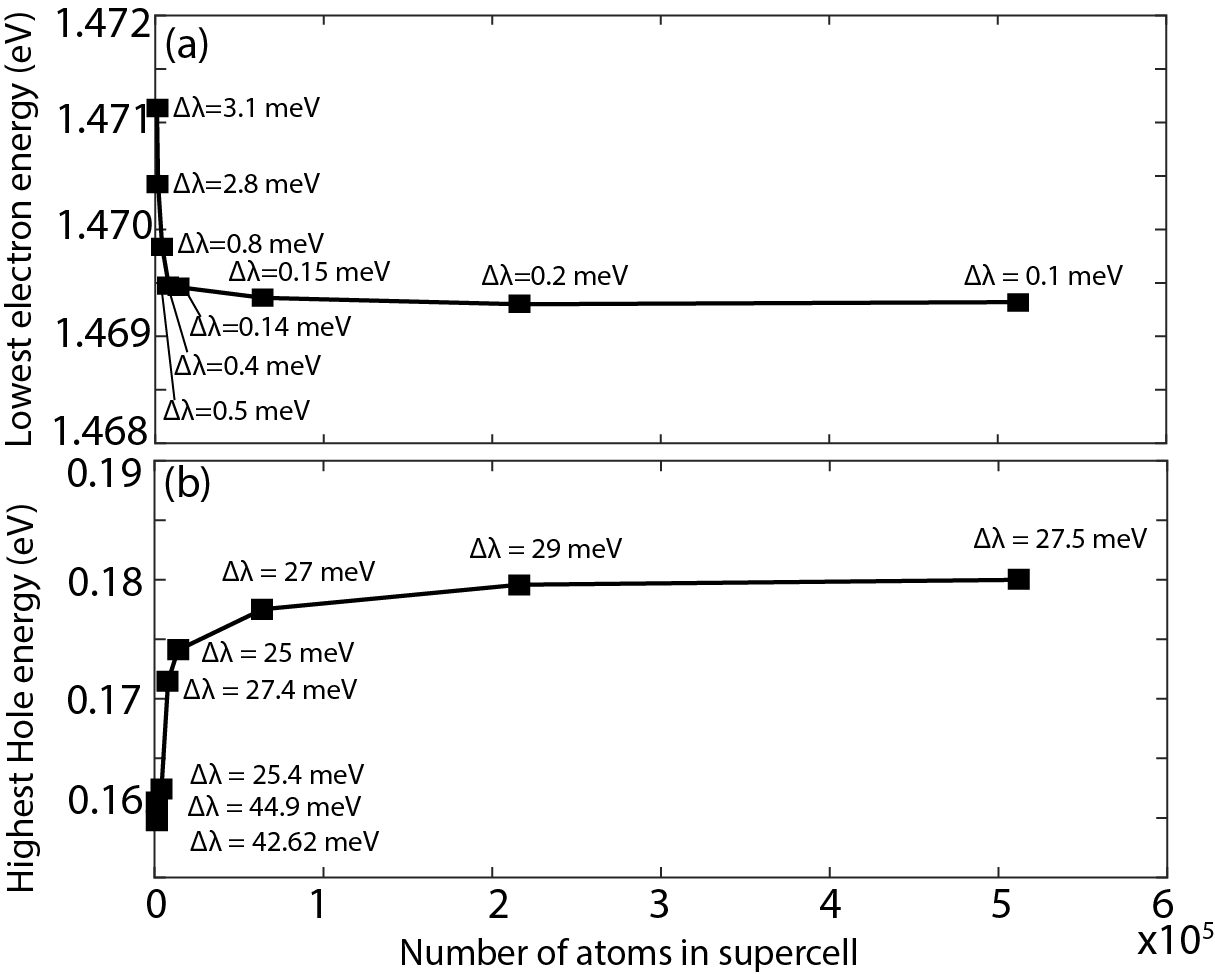}
\caption{(a) The lowest electron energy is plotted as a function of the number of atoms in the simulated \GaBiAs supercell containing 3.125\% Bi and strained on GaAs substrate. For each supercell size, the broadening ($\Delta \lambda$) of electron energy is computed by performing simulations for five different random configurations of Bi atoms. (b) Same as (a) for the highest hole energy.}
\label{fig:FigS1}
\end{figure*}

\begin{figure*}
\includegraphics[scale=0.4]{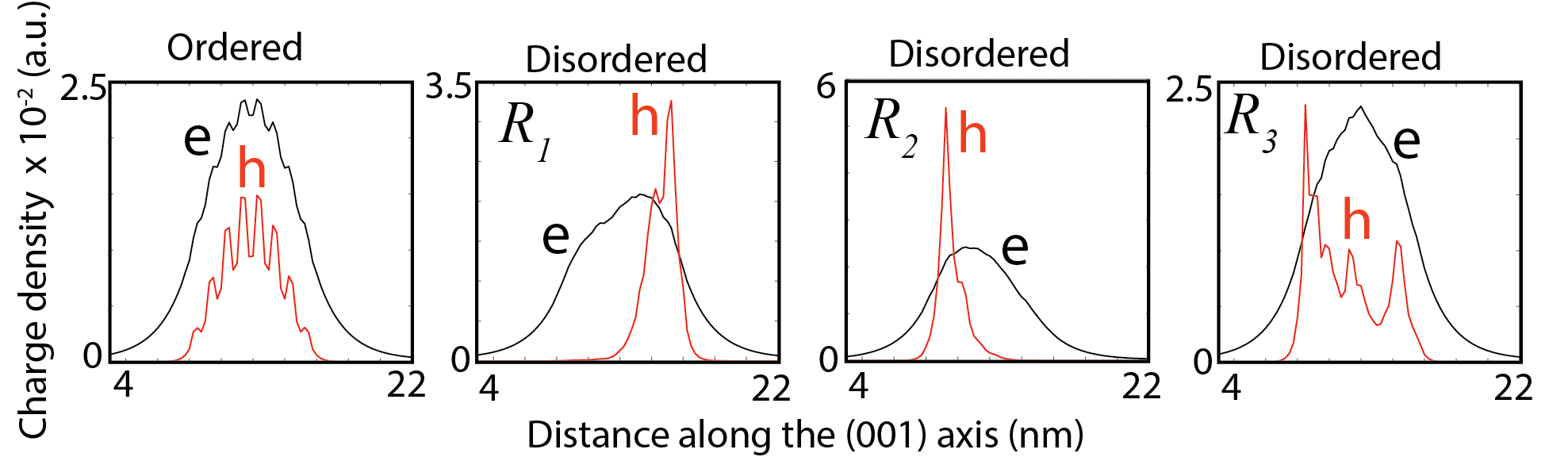}
\caption{Plots of the lowest electron (e) and the highest hole (h) charge densities ($\vert \Psi \vert^2$) are shown for single QW devices and for both ordered and disordered placement of Bi atoms. To highlight the impact of disorder, we plot three different random placements of Bi atoms ($R_1, R_2, R_3$). The electron states are weakly perturbed by disorder, whereas the hole states are high confined and clearly show significant impact from Bi pairs and clustering.}
\label{fig:FigS2}
\end{figure*}

\begin{figure*}
\includegraphics[scale=0.4]{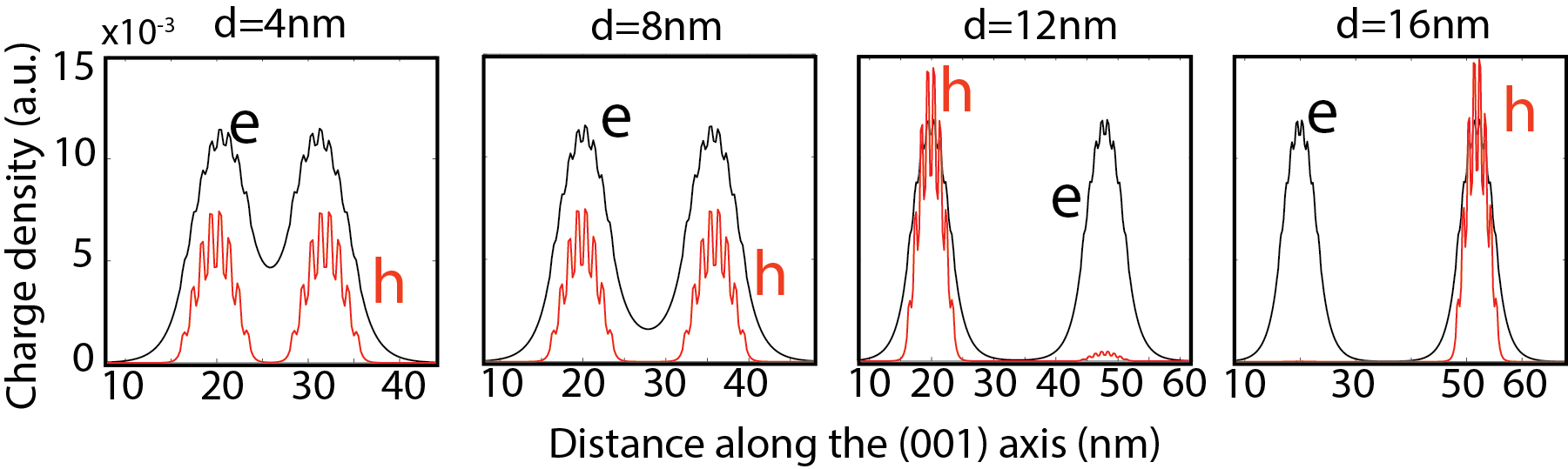}
\caption{Plots of the lowest electron (e) and the highest hole (h) charge densities ($\vert \Psi \vert^2$) are shown as a function of distance along the (001) axis are shown for a few selected separations $d$ for double QW structures. For small separations between the QWs, strong effect of coupling leads to bonding-like hybridized states spread over all the QWs as well as on the GaAs spacer in-between them. For large separations, the coupling effect becomes weak for electron states and completely vanishes for the hole states.}
\label{fig:FigS3}
\end{figure*}

\begin{figure*}
\includegraphics[scale=0.5]{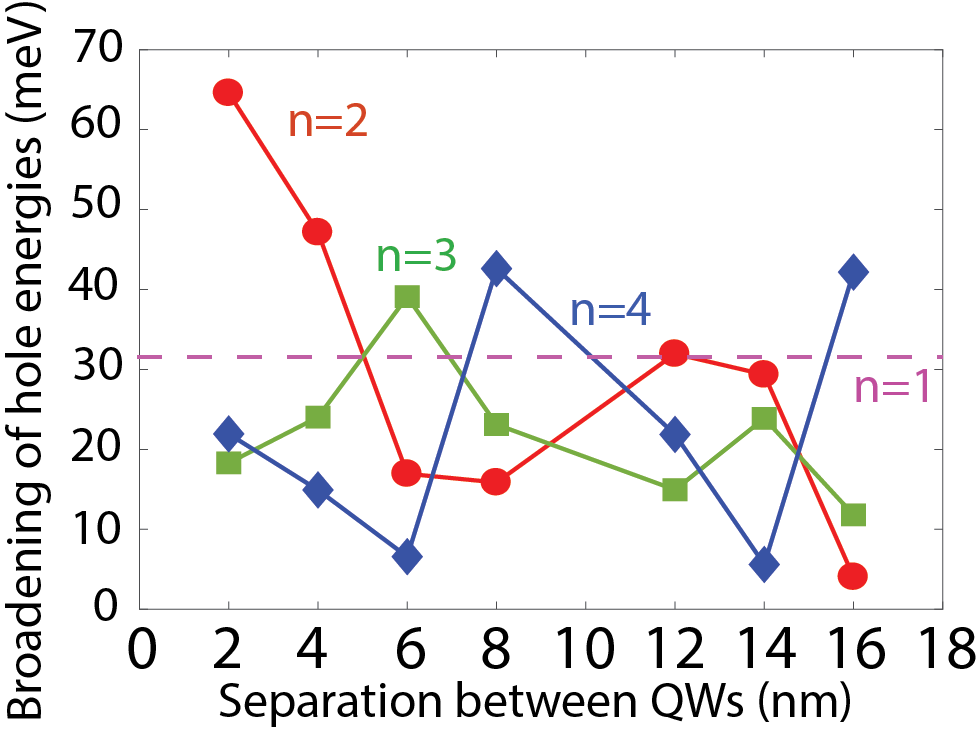}
\caption{Broadening of the highest hole energy levels is plotted for various QW devices ($n$=1, 2, 3, and 4) as a function of separation $d$ between the QWs. The broadening is computed from the difference between minimum and maximum of hole energies computed from five different random placement of Bi atoms in the QW regions.}
\label{fig:FigS4}
\end{figure*}

\begin{figure*}
\includegraphics[scale=0.4]{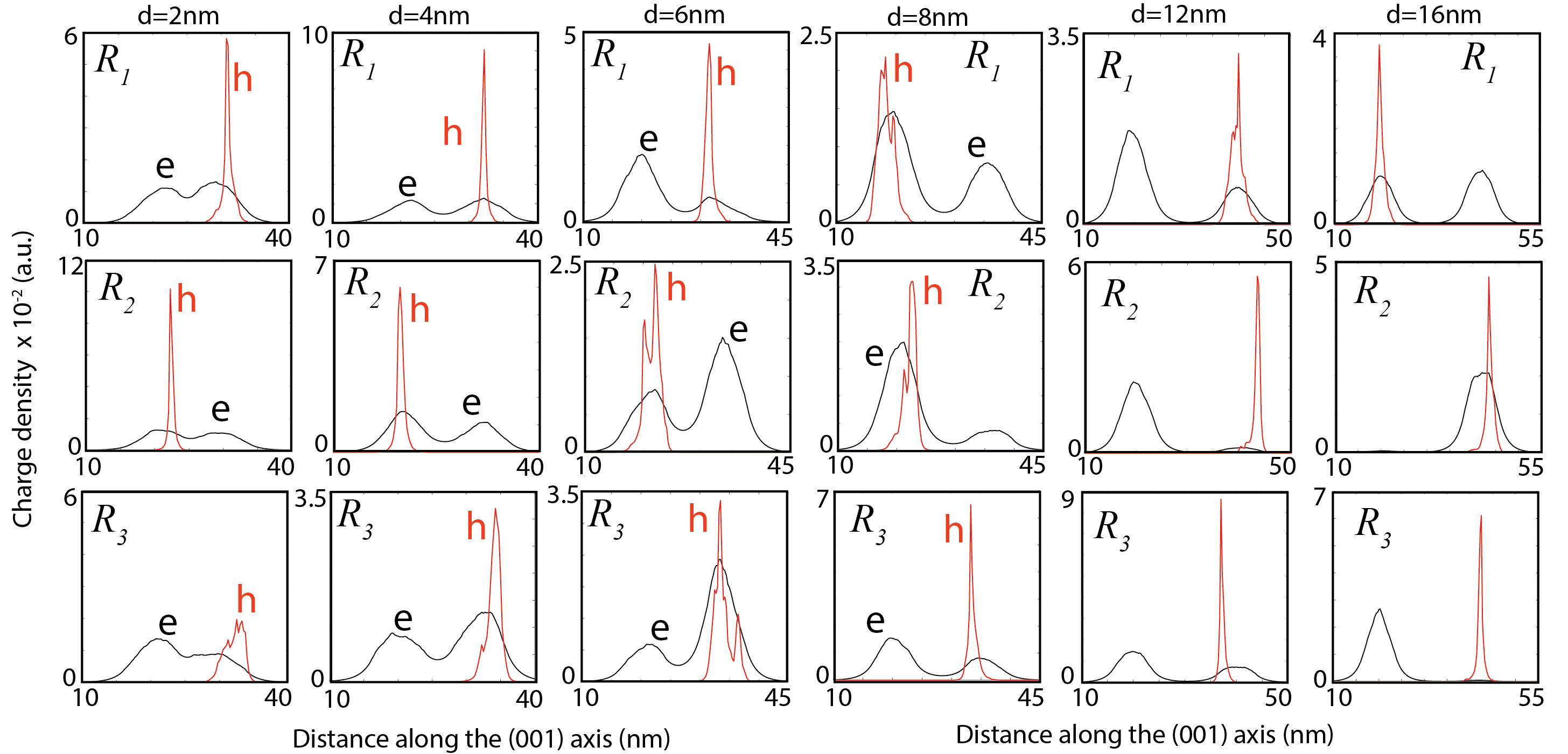}
\caption{Plots of the lowest electron (e) and the highest hole (h) state charge distributions ($\vert \Psi \vert^2$) are shown for double QW structures ($n$=2) as a function of the distance along the (001) axis for various QW separations ($d$). For each device, we plot three random distributions ($R_1, R_2, R_3$) of Bi atoms to show the character of alloy disorder on electron and hole states.}
\label{fig:FigS5}
\end{figure*}

\begin{figure*}
\includegraphics[scale=0.4]{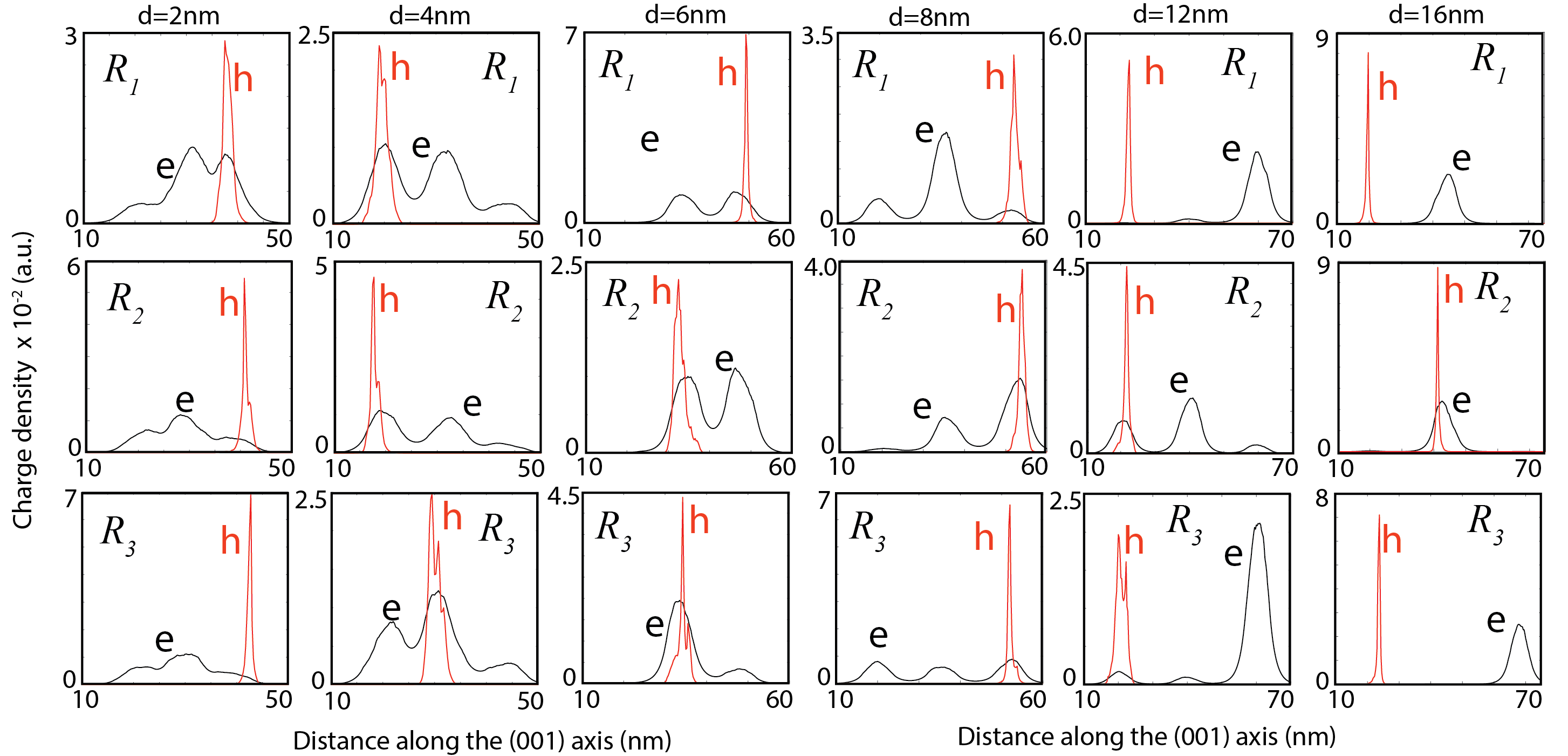}
\caption{Plots of the lowest electron (e) and the highest hole (h) state charge distributions ($\vert \Psi \vert^2$) are shown for three QW structures ($n$=3) as a function of the distance along the (001) axis for various QW separations ($d$). For each device, we plot three random distributions ($R_1, R_2, R_3$) of Bi atoms to show the character of alloy disorder on electron and hole states.}
\label{fig:FigS6}
\end{figure*}

\begin{figure*}
\includegraphics[scale=0.4]{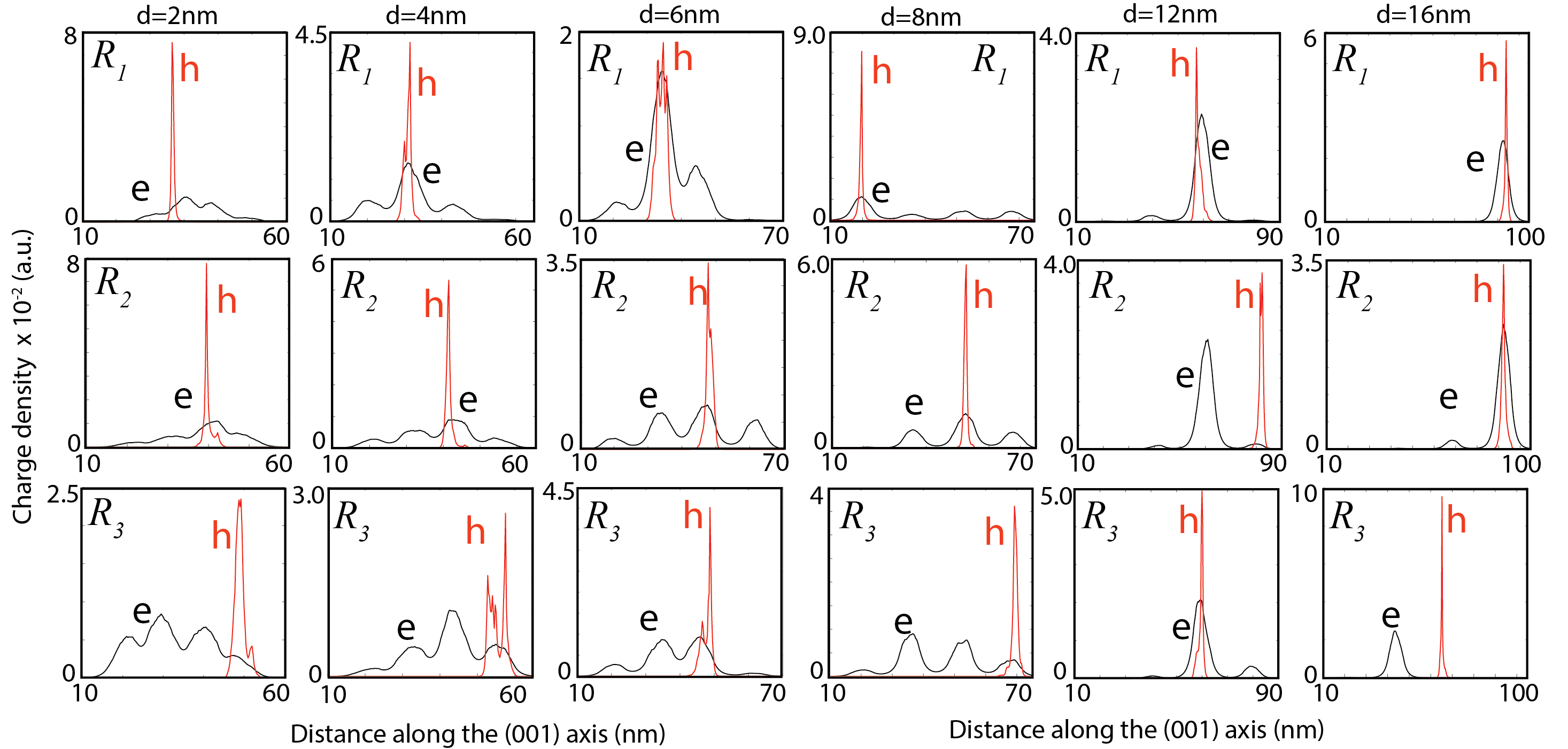}
\caption{Plots of the lowest electron (e) and the highest hole (h) state charge distributions ($\vert \Psi \vert^2$) are shown for four QW structures ($n$=4) as a function of the distance along the (001) axis for various QW separations ($d$). For each device, we plot three random distributions ($R_1, R_2, R_3$) of Bi atoms to show the character of alloy disorder on electron and hole states.}
\label{fig:FigS7}
\end{figure*}

\begin{figure*}
\includegraphics[scale=0.4]{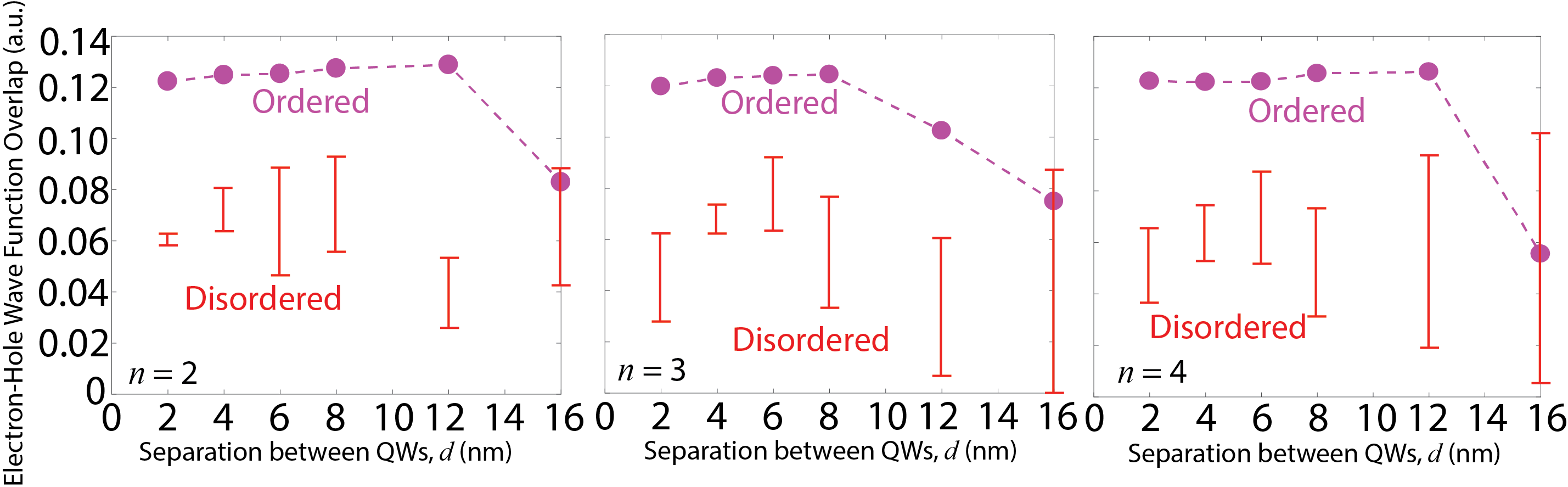}
\caption{Plots of overlap between the lowest electron and the highest hole wave functions are shown for both ordered and disordered QW structures, and for $n$=2, 3, and 4.}
\label{fig:FigS8}
\end{figure*}

\begin{figure*}
\includegraphics[scale=0.4]{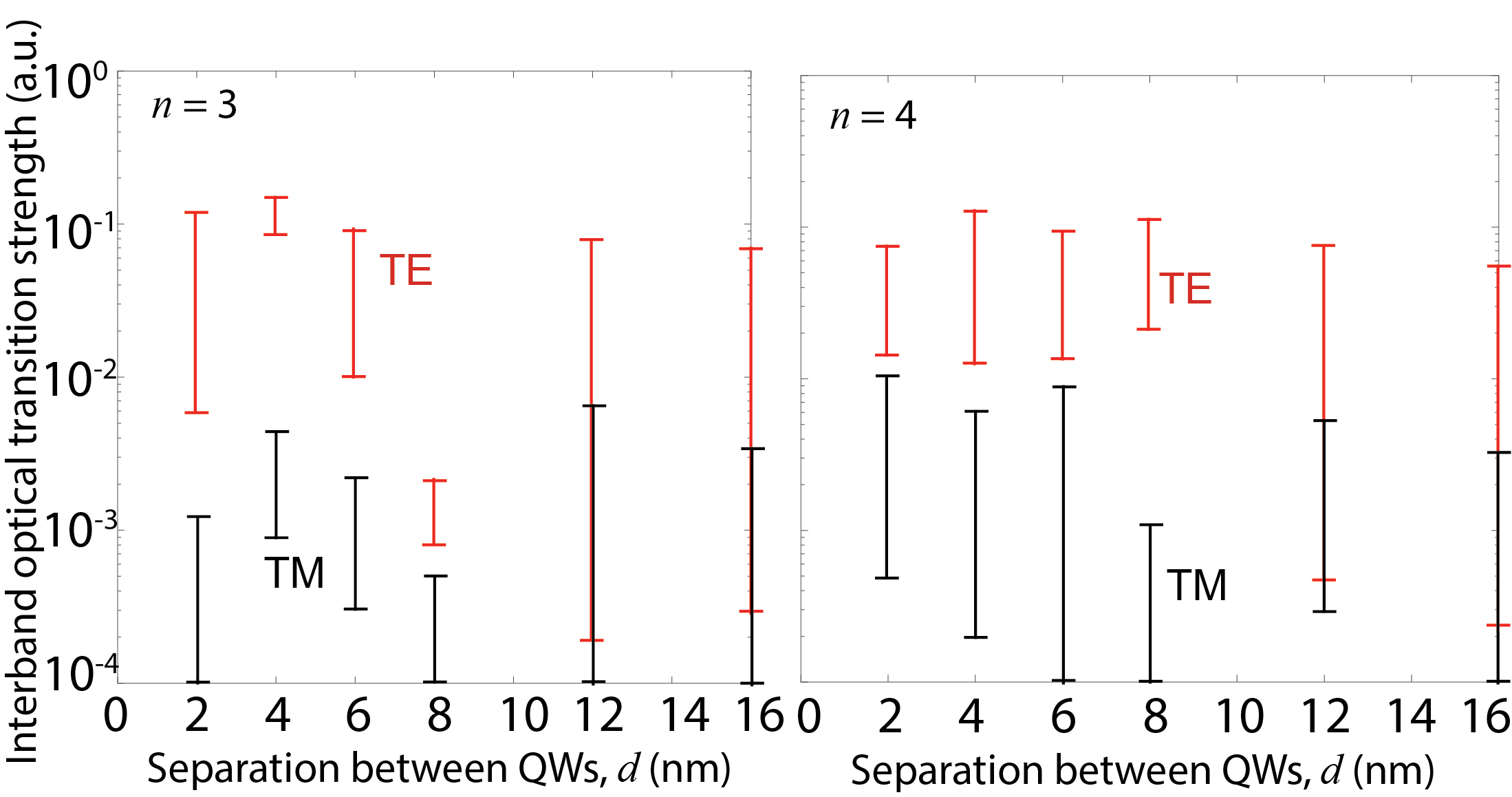}
\caption{Polarisation-resolved TE and TM inter-band optical transition strengths are plotted as a function of QW separations for triple and four QW structures. }
\label{fig:FigS9}
\end{figure*}

\begin{figure*}
\includegraphics[scale=0.5]{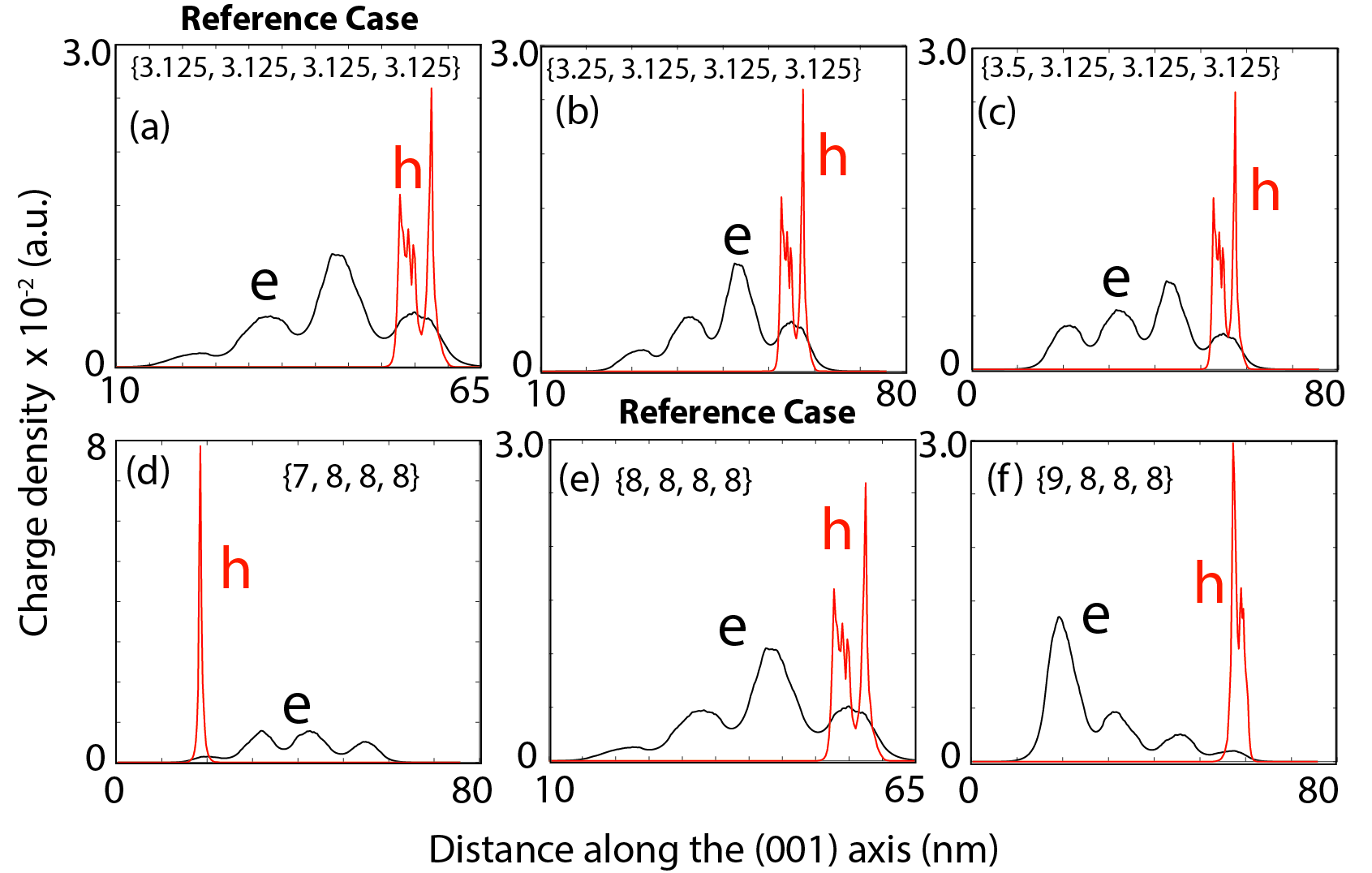}
\caption{(a-c) The Bi fraction of a four QW device ($n$=4) is varied as indicated by the labels \{$x_1,x_2,x_3,x_4$\}, where $x_1$,$x_2$, $x_3$ and $x_4$ are the Bi fractions of the lowest to the topmost QW regions, respectively. The widths $w$ of all QWs are 8 nm and the separation between the QWs is 4 nm. (d-f) The width of the lowest QW in a four QW structure ($n$=4) is varied as indicated by the labels \{$w_1,w_2,w_3, w_4$\}, where $w_1$,$w_2$, $w_3$, and $w_4$ are the widths of the QWs from the lowest to the topmost, respectively. The Bi fraction $x$ of all QWs is 3.125\% and the separation between the QWs is 4 nm.}
\label{fig:FigS10}
\end{figure*}

\bibliographystyle{apsrev4-1}
%

\end{document}